\documentclass[11pt,a4paper]{article}
\pdfoutput=1
\DeclareUnicodeCharacter{2212}{-}

\usepackage{jheppub}
\usepackage{bm}
\usepackage{mathtools}
\usepackage{booktabs}
\usepackage{multirow}
\usepackage{array}
\usepackage{siunitx}
\usepackage{enumitem}
\usepackage{microtype}
\usepackage{float}
\usepackage{upgreek}
\usepackage[T1]{fontenc}
\usepackage{xcolor} 
\hypersetup{
  colorlinks=true,
  linkcolor=blue,
  citecolor=blue,
  urlcolor=blue
}

\graphicspath{{plots/}{./}}
\allowdisplaybreaks

\newcommand{\sqrtsFCC}{240~\mathrm{GeV}}
\newcommand{\lumiFCC}{10.8~\mathrm{ab}^{-1}}
\newcommand{\mCMS}{146~\mathrm{GeV}}
\newcommand{\hCBcms}{1.0 \times 10^{-2}}          
\newcommand{\hSSSFcms}{1.7 \times 10^{-3}}    
\newcommand{\haecms}{1.46\times10^{-2}}
\newcommand{\hmuoniumcms}{5.98\times10^{-2}}
\newcommand{\hLEPcms}{7.08\times10^{-2}}
\newcommand{\hamucms}{6.91\times10^{-1}}

\newcommand{\mx}{m_X}

\newcommand{\BR}{\operatorname{BR}}

\newcommand{\pt}{p_T}
\newcommand{\Lag}{\mathcal{L}}
\newcommand{\GeV}{\mathrm{GeV}}

\newcommand{\draftfigure}[3]{%
  \IfFileExists{#1}{%
    \includegraphics[width=#2]{#1}%
  }{%
    \fbox{%
      \parbox[c][0.26\textheight][c]
      {\dimexpr#2-2\fboxsep-2\fboxrule\relax}{%
        \centering
        \textbf{Figure to be inserted}\\[4pt]
        \texttt{\detokenize{#1}}\\[6pt]
        #3
      }%
    }%
  }%
}

\title{\boldmath
Charge-resolved four-lepton probes of an
$e\mu$-flavour-violating scalar at FCC-ee
}

\author[b]{Kuldeep Deka,}
\author[a]{Bharti Rawat,}
\author[a]{and Tripurari Srivastava}

\affiliation[a]{Department of Physics and Astrophysics,\\
University of Delhi, Delhi 110007, India}

\affiliation[b]{New York University Abu Dhabi,\\
PO Box 129188, Saadiyat Island, Abu Dhabi, United Arab Emirates\\
and\\
Ruđer Bošković Institute, 10000 Zagreb, Croatia}

\emailAdd{kuldeepdeka.physics@gmail.com}
\emailAdd{bhartirawat87@gmail.com}
\emailAdd{tripurarisri022@gmail.com}

\abstract{
We study a real neutral scalar $X$ with a single flavour-off-diagonal
coupling $h_{e\mu}=h_{\mu e}\equiv h$. The coupling induces associated
production, $e^+e^-\to e^\pm\mu^\mp X$, and the decay
$X\to e^\pm\mu^\mp$. The resulting four-lepton final state separates into
the charge-balanced channel $e^+e^-\mu^+\mu^-$ and the same-sign
same-flavour channels $e^\pm e^\pm\mu^\mp\mu^\mp$. The charge-balanced
channel receives a large prompt continuum background, while the same-sign category has a small estimated background under our benchmark assumptions. We analyse the two channels at FCC-ee for
$\sqrt{s}=\sqrtsFCC$ and $\mathcal{L}=\lumiFCC$, including a detector
simulation. The collider sensitivities are compared with the one-coupling
constraints from the electron and muon magnetic moments,
muonium--antimuonium conversion, and LEP dilepton data. At $m_X=\mCMS$,
motivated by the largest excess in the CMS $e\mu$ resonance search, the
expected 95\% CL reach of the charge-balanced channel, including a 1\%
background uncertainty, is $|h|\simeq\hCBcms$, while the same-sign channel
reaches $|h|\simeq\hSSSFcms$, compared with the cesium-based electron-$g-2$ bound $|h|<\haecms$. Both channels thus probe couplings below the present bound. The two signal rates are equal before selection; the difference in reach follows from the different backgrounds.
}

\keywords{charged-lepton flavour violation, neutral scalar, FCC-ee,
future lepton colliders}

\begin{document}
\maketitle
\flushbottom

\section{Introduction}
\label{sec:introduction}
The discovery of the Higgs boson by the ATLAS and CMS
Collaborations in 2012 completed the experimental
identification of the Standard Model (SM) particle
spectrum \cite{ATLAS:2012yve,CMS:2012qbp}.
Despite its success, the SM does not explain neutrino
masses and mixing, the nature of dark matter, or the
observed baryon asymmetry of the Universe. These
limitations motivate extensions of the SM, including
models with additional scalar degrees of freedom \cite{Branco:2011iw,Harnik:2012pb}.
In such models, the Yukawa interactions of neutral
scalars need not be flavour diagonal, allowing
transitions between charged leptons of different
flavours \cite{Harnik:2012pb,Dev:2017ftk}. 

Charged-lepton flavour violation (CLFV) is a sensitive
probe of these interactions \cite{deGouvea:2013zba,Lindner:2016bgg,
Calibbi:2017uvl,Ardu:2022sbt}. In the SM extended to
include light-neutrino masses, CLFV processes are
suppressed far below present experimental sensitivities \cite{Calibbi:2017uvl,deGouvea:2013zba}.
An observable CLFV signal would therefore establish
the existence of new physics. Neutral scalars with
off-diagonal lepton Yukawa couplings provide a simple
realisation of such interactions \cite{Harnik:2012pb,Dev:2017ftk}. Low-energy observables,
including $\mu\to e\gamma$, $\mu\to 3e$, and
muonium–antimuonium conversion \cite{MEGII:2023ltw,SINDRUM:1987nra,Willmann:1998gd}, constrain different
combinations of these couplings \cite{Calibbi:2017uvl,Willmann:1998gd,
Hou:1995dg,Dev:2017ftk}. In the single off-diagonal $e\mu$ coupling scenario,
the electron and muon anomalous magnetic moments,
muonium--antimuonium conversion, and LEP dilepton
measurements provide particularly relevant constraints
\cite{Dev:2017ftk,Hou:1995dg,Willmann:1998gd,
Conlin:2020veq}. 
Direct searches at
colliders offer complementary sensitivity, particularly
when the new scalar can be produced on shell \cite{Dev:2017ftk}.

CMS has searched for a neutral resonance decaying into
$e^\pm\mu^\mp$ in the mass range
$110$--$160~\mathrm{GeV}$ using the full Run-II
data set \cite{CMS:2023pte}. The largest excess was observed near
$146~\mathrm{GeV}$, with local and global
significances of $3.8\sigma$ and $2.8\sigma$,
respectively \cite{CMS:2023pte}. Although the
excess is not statistically conclusive, it provides
a useful benchmark for searches involving an
off-diagonal electron-muon coupling. Scalar
interpretations of this result have been considered
in Refs.~\cite{Primulando:2023ugc,Afik:2023vyl,
Koivunen:2023led,Gao:2026qnt}. 
Producing the scalar at the LHC at the rate suggested by the excess
requires couplings to quarks or gluons, or mixing with the Higgs boson,
which are absent in the minimal scenario considered here. We therefore
use $m_X=146~\mathrm{GeV}$ as a benchmark mass for the lepton-collider
analysis.

Future electron-positron colliders provide a clean
environment for investigating such interactions \cite{Dev:2017ftk,
Li:2018cod,Calibbi:2021pyh}.
Their well-defined initial states and relatively
low hadronic backgrounds are advantageous for
multi-lepton searches \cite{Kabachenko:1997aw}. The Future Circular Collider
in its electron-positron configuration (FCC-ee),
with its anticipated high luminosity, offers an
opportunity to probe flavour-violating scalar
interactions through associated production \cite{TLEPDesignStudyWorkingGroup:2013myl}.
Lepton-flavour-violating Higgs decays at FCC-ee~\cite{Sriling:2026vmx} and
same-sign dilepton probes of charged-lepton flavour violation at
electron--photon colliders~\cite{Zhang:2026dui} have also been studied recently. 

For a neutral scalar $X$ coupled to electrons and muons,
the relevant on-shell process is $e^+e^-\to e^\pm\mu^\mp X$.
Dev et al.~\cite{Dev:2017ftk} studied this process and estimated
the background to the associated $e\mu$ pair from $Zh$ events
with a misidentified lepton. Their sensitivity projection used
a ten-signal-event criterion and benchmark fractions of
reconstructible scalar decays.

We reconstruct $X\to e^\pm\mu^\mp$ and analyse the resulting
four-lepton events in two charge categories. The charge-balanced
(CB) final state is $e^+e^-\mu^+\mu^-$, and the same-sign same-flavour
(SSSF) final states are $e^\pm e^\pm\mu^\mp\mu^\mp$. For the
symmetric coupling considered here, the two signal rates are equal
before selection. The CB channel receives an irreducible SM
$e^+e^-\mu^+\mu^-$ continuum, so its sensitivity requires an
explicit treatment of this background. In the SSSF channel, the
estimated contributions from four-$\tau$ and $WW\tau\tau$ production along with charge
misidentification are small after selection under our benchmark
assumptions. We include detector effects, evaluate the FCC-ee
sensitivity separately for each channel, and compare the projected
reach with existing bounds in the $(m_X,|h_{e\mu}|)$ plane.

The paper is organised as follows.
Section~\ref{sec:model} introduces the scalar
interaction and the charge decomposition of
the signal. Section~\ref{sec:constraints}
discusses the constraints in the
$(m_X,|h_{e\mu}|)$ plane.
The collider analysis is presented in
Sec.~\ref{sec:collider}, followed by the
projected sensitivities in
Sec.~\ref{sec:results}.

\section{Scalar interaction and charge structure}
\label{sec:model}
We first specify the scalar interactions relevant to our analysis and identify the charge configurations arising from its lepton-flavour-violating production and decay. We work in a minimal scenario with a single off-diagonal electron-muon coupling, allowing the collider signatures and low-energy constraints to be described within the same parameter space.
\subsection{One-coupling setup}

We add one real CP-even neutral scalar $X$ and write its couplings to charged
leptons in the mass basis as
\begin{equation}
  -\Lag_Y
  = X\,\overline{\ell_\alpha}
  \left(h_{\alpha\beta}P_R+h_{\beta\alpha}^{*}P_L\right)
  \ell_\beta .
  \label{eq:general-yukawa}
\end{equation}
The main analysis uses
\begin{equation}
  h_{e\mu}=h_{\mu e}\equiv h\in\mathbb{R},
  \qquad
  h_{ee}=h_{\mu\mu}=h_{\tau\tau}=h_{e\tau}=h_{\mu\tau}=0.
  \label{eq:one-coupling-slice}
\end{equation}
In this limit,
\begin{equation}
  -\Lag_Y=hX\left(\overline e\mu+\overline\mu e\right).
  \label{eq:symmetric-interaction}
\end{equation}
We also set the mixing of $X$ with the 125 GeV Higgs to zero and neglect
couplings to quarks and gluons. The parameter space is therefore the
two-dimensional plane
\begin{equation}
  (\mx,|h|).
\end{equation}
We restrict the analysis to this two-parameter slice so that the low-energy
and collider observables depend on the same coupling without introducing
additional Yukawa products.

Neglecting the charged-lepton masses, the charge-summed width is
\begin{equation}
  \Gamma\!\left(X\to e^\pm\mu^\mp\right)
  =\frac{\mx}{8\pi}
  \left(|h_{e\mu}|^2+|h_{\mu e}|^2\right)
  =\frac{h^2\mx}{4\pi}.
  \label{eq:X-width}
\end{equation}
Since no other decay channel is present in Eq.~\eqref{eq:one-coupling-slice},
\begin{equation}
  \BR(X\to e^\pm\mu^\mp)=1.
  \label{eq:BR-one}
\end{equation}
The scalar is narrow in the coupling range probed by the collider analysis.
The on-shell cross section therefore scales as
\begin{equation}
  \sigma\bigl(e^+e^-\to e\mu X\to e\mu e\mu\bigr)\propto h^2,
  \label{eq:signal-h-scaling}
\end{equation}
provided the width is changed consistently with the coupling.

\subsection{Charge decomposition of the signal}

Let
\begin{align}
  P_+&=\sigma(e^+e^-\to e^+\mu^-X),
  &P_-&=\sigma(e^+e^-\to e^-\mu^+X),\\
  B_+&=\BR(X\to e^+\mu^-),
  &B_-&=\BR(X\to e^-\mu^+).
\end{align}
The two charge categories have cross sections
\begin{align}
  \sigma_{\rm CB}&=P_+B_-+P_-B_+,
  \label{eq:sigma-CB}\\
  \sigma_{\rm SSSF}&=P_+B_+ +P_-B_-.
  \label{eq:sigma-SSSF}
\end{align}
For a real scalar with the symmetric coupling in
Eq.~\eqref{eq:one-coupling-slice},
\begin{equation}
  P_+=P_-,\qquad B_+=B_-=\frac{1}{2},
\end{equation}
up to charge-asymmetric acceptance effects. Hence
\begin{equation}
  \sigma_{\rm CB}\simeq \sigma_{\rm SSSF}.
  \label{eq:equal-charge-rates}
\end{equation}
This equality is a useful check of the event generation before cuts.

\begin{figure}[t]
  \centering
  \includegraphics[width=8cm,height=4.5cm]{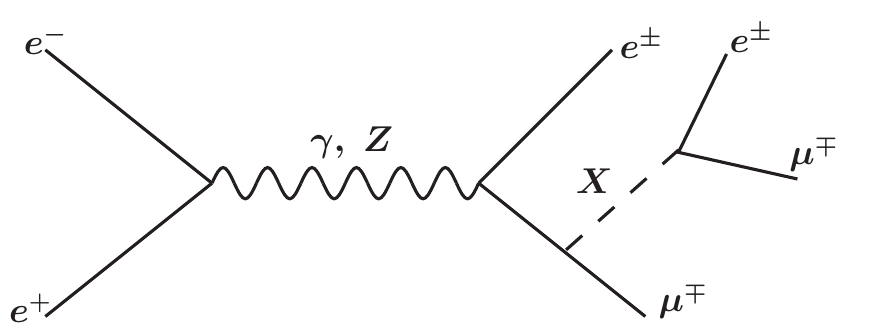}
  \includegraphics[width=6.8cm,height=5cm]{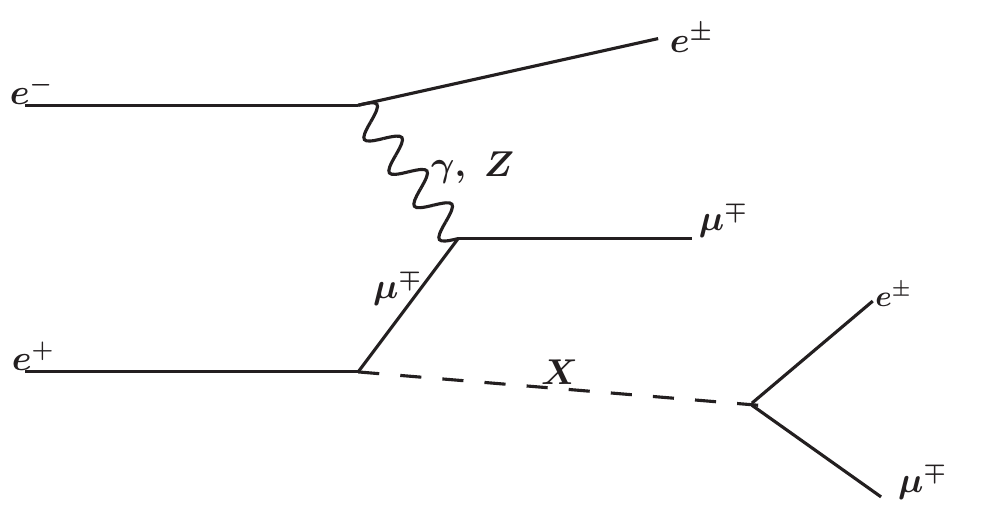}
  \caption{Representative diagrams for associated production of the
  flavour-violating scalar. Signal events are generated in the
  narrow-width approximation for $X$.}
  \label{fig:signal-diagrams}
\end{figure}

\section{Present constraints in the one-coupling plane}
\label{sec:constraints}
We first examine the experimental constraints relevant to the one-coupling scenario introduced in Sec.~\ref{sec:model}. With all other scalar Yukawa couplings set to zero, the observables directly constraining $h$ are muonium–antimuonium conversion, the electron and muon anomalous magnetic moments, and the LEP measurement of $e^+e^-\to\mu^+\mu^-$. These constraints are evaluated in the $(m_X,|h|)$ plane, using the exact one-loop expressions for the magnetic moments, and subsequently compared with the projected FCC-ee sensitivity.

\subsection{Muonium--antimuonium conversion}

Tree-level $X$ exchange induces conversion between muonium and
antimuonium. The transition matrix element is \cite{Hou:1995dg}
\begin{equation}
  |\mathcal{M}|
  \equiv\bigl|\langle\overline{M}|\mathcal{H}_{\rm eff}|M\rangle\bigr|
  =\frac{\alpha_{\rm em}^{3}\mu_r^3}{\pi\mx^2}|h|^2,
  \qquad
  \mu_r=\frac{m_em_\mu}{m_e+m_\mu},
  \label{eq:muonium-deltaM}
\end{equation}
corresponding to a zero-field mass splitting $|\Delta M|=2|\mathcal{M}|$ between the
eigenstates. For a scalar--scalar operator, the probability measured in a
magnetic field is
\begin{equation}
  P_B(M\to\overline M)
  =S_B\,
  \frac{2|\mathcal{M}|^2}{\Gamma_\mu^2+4|\mathcal{M}|^2}.
  \label{eq:muonium-probability}
\end{equation}
The MACS result gives
\begin{equation}
  P_B(M\to\overline M)<8.3\times10^{-11}
  \qquad (90\%~\mathrm{C.L.})
  \label{eq:muonium-limit}
\end{equation}
for a field of $0.1$ T \cite{Willmann:1998gd}. The magnetic field lifts the
degeneracy of the muonium and antimuonium states and suppresses the
conversion by a factor $S_B$ that depends on the Lorentz structure of the
interaction. Exchange of $X$ with the scalar coupling of
Eq.~\eqref{eq:symmetric-interaction} generates an operator of
scalar--scalar type, for which $S_B=0.50$ at $0.1$ T
\cite{Willmann:1998gd}. At $\mx=\mCMS$ this gives
\begin{equation}
  |h|<\hmuoniumcms
  \qquad (90\%~\mathrm{C.L.}).
  \label{eq:muonium-146}
\end{equation}
\subsection{Electron and muon magnetic moments}

For an external lepton $\ell$ and an internal lepton $f$, the exact one-loop
contribution from a real scalar is \cite{Lindner:2016bgg}
\begin{equation}
  \Delta a_\ell^X
  =\frac{|h|^2}{8\pi^2}\frac{m_\ell^2}{\mx^2}
  \int_0^1 dx\,
  \frac{x^2(1-x+\epsilon)}
  {(1-x)(1-x\lambda^2)+x\epsilon^2\lambda^2},
  \label{eq:exact-gminus2}
\end{equation}
where
\begin{equation}
  \epsilon=\frac{m_f}{m_\ell},
  \qquad
  \lambda=\frac{m_\ell}{\mx}.
\end{equation}
For the electron magnetic moment, $\ell=e$ and $f=\mu$. In the heavy-scalar
limit this reduces to
\begin{equation}
  \Delta a_e^X\simeq
  \frac{|h|^2m_em_\mu}{16\pi^2\mx^2}
  \left[2\ln\!\left(\frac{\mx^2}{m_\mu^2}\right)-3\right].
  \label{eq:ae-heavy}
\end{equation}
We use the electron magnetic moment measurement of
Ref.~\cite{Fan:2022eto}, the cesium determination of the fine-structure
constant \cite{Parker:2018vye}, and the revised tenth-order QED coefficient
\cite{Aoyama:2024aly}. These inputs give
\begin{equation}
  \Delta a_e^{\rm Cs}=(-9.97\pm2.61)\times10^{-13}.
  \label{eq:ae-Cs-residual}
\end{equation}
The scalar contribution is positive. We therefore obtain the upper limit from
the likelihood ratio with respect to the physical point $\Delta a_e^X=0$,
\begin{equation}
  \Delta\chi_e^2(h)
  =\frac{(\Delta a_e^{\rm Cs}-\Delta a_e^X(h))^2
  -(\Delta a_e^{\rm Cs})^2}{\sigma_e^2},
  \qquad \Delta\chi_e^2=2.71.
  \label{eq:ae-likelihood}
\end{equation}
This is the strongest direct one-coupling bound in the mass range shown in
our main plot. At $\mx=\mCMS$,
\begin{equation}
  |h|<\haecms
  \qquad (95\%~\mathrm{C.L.},~\text{Cs input}).
  \label{eq:ae-146}
\end{equation}
The independent rubidium determination of $\alpha$~\cite{Morel:2020dww} gives a weaker limit for
a positive scalar contribution. The main figure uses the stronger cesium curve.
This bound is strengthened by the negative residual in
Eq.~\eqref{eq:ae-Cs-residual}; for a vanishing residual with the same
uncertainty, the limit at $\mx=\mCMS$ would be weaker by a factor of about two.

The final Fermilab result and the 2025 theory average give
\begin{equation}
  \Delta a_\mu
  \equiv a_\mu^{\rm exp}-a_\mu^{\rm SM}
  =(38\pm63)\times10^{-11},
  \label{eq:amu-current}
\end{equation}
which is consistent with zero \cite{Muong-2:2026qnz,Aliberti:2025beg,ParticleDataGroup:2026aaa}.
For $h_{e\mu}$, the internal lepton in Eq.~\eqref{eq:exact-gminus2} is the
electron. The contribution is therefore much smaller than the familiar
$h_{\mu\tau}$ case, where an internal tau provides a large chirality
enhancement. Requiring the positive scalar contribution to lie below the
one-sided 95\% upper limit gives
\begin{equation}
  |h|<\hamucms
  \qquad \text{at}\qquad \mx=\mCMS.
  \label{eq:amu-146}
\end{equation}
Thus, the new near-Standard-Model value of $a_\mu$ changes the old preferred
region into an upper bound, but it is not competitive in the $e\mu$ coupling
plane.

\subsection{LEP dilepton data}

The same coupling contributes to
\begin{equation}
  e^+e^-\to\mu^+\mu^-
\end{equation}
through $t$-channel scalar exchange. In the heavy-mediator limit, the LEP
constraint used in Ref.~\cite{Dev:2017ftk} can be written as
\begin{equation}
  \frac{|h|^2}{\mx^2}<1.6\times10^{-7}~\GeV^{-2}.
  \label{eq:lep-heavy-limit}
\end{equation}
For masses below the LEP centre-of-mass energy, a contact-interaction
extrapolation is not adequate. We use the same angle-averaged finite
propagator prescription,
\begin{equation}
  |h|_{\rm LEP}
  \simeq
  \left[
  1.6\times10^{-7}
  \left(\mx^2+\frac{s_{\rm LEP}}{4}\right)
  \right]^{1/2},
  \qquad \sqrt{s_{\rm LEP}}=200~\GeV,
  \label{eq:lep-propagator}
\end{equation}
based on the DELPHI dilepton analysis \cite{DELPHI:2005wxt}. This approximation follows Ref.~\cite{Dev:2017ftk}. At $\mx=\mCMS$ it gives
\begin{equation}
  |h|<\hLEPcms.
  \label{eq:lep-146}
\end{equation}

\subsection{Constraints requiring additional couplings}

The latest limits on $\mu\to e\gamma$, $\mu\to3e$, and
$\mu N\to eN$ are important in models with more than one nonzero coupling.
They do not give independent curves in the strict one-coupling plane of
Eq.~\eqref{eq:one-coupling-slice}. At leading order,
\begin{align}
  \mu\to e\gamma &: \quad
  h_{e\tau}h_{\mu\tau},~h_{ee}h_{e\mu},~h_{\mu\mu}h_{e\mu},\nonumber\\
  \mu\to3e &: \quad h_{ee}h_{e\mu},\nonumber\\
  \mu N\to eN &: \quad h_{e\mu}\times
  (\text{quark/gluon coupling or Higgs mixing}).
  \label{eq:product-coupling-observables}
\end{align}
These amplitudes vanish at the order considered when the
additional couplings are set to zero. Their experimental
limits therefore do not provide independent bounds on $h$
in this setup.

\begin{table}[t]
  \centering
  \caption{Direct one-coupling constraints at $\mx=146~\GeV$ in the setup of
  Eq.~\eqref{eq:one-coupling-slice}. The electron and muon magnetic-moment
  entries are one-sided 95\% C.L. bounds on a positive scalar contribution;
  the muonium entry is at 90\% C.L., as published.}
  \label{tab:constraints-146}
  \begin{tabular}{lc}
    \toprule
    Observable & Upper bound on $|h|$ \\
    \midrule
    $(g-2)_e$ with Cs $\alpha$ & $\haecms$ \\
    Muonium--antimuonium conversion & $\hmuoniumcms$ \\
    LEP $e^+e^-\to\mu^+\mu^-$, recast & $\hLEPcms$ \\
    $(g-2)_\mu$ with the current SM comparison & $\hamucms$ \\
    \bottomrule
  \end{tabular}
\end{table}

\section{Charge-resolved analysis at FCC-ee}
\label{sec:collider}
We now investigate the collider phenomenology of the flavour-violating scalar at FCC-ee.
\par We study associated production of \(X\) with an electron–muon pair,\footnote{We extend the associated-production study of
Ref.~\cite{Dev:2017ftk} by reconstructing the scalar decay after
detector simulation, fitting the CB invariant-mass spectrum, and
estimating the residual SSSF backgrounds separately. We also update
the electron and muon magnetic-moment inputs, recalculate the
muonium--antimuonium conversion bound for the scalar interaction,
and include the LEP dilepton recast in the comparison
(Sec.~\ref{sec:constraints}).}
\begin{equation}
	e^+e^-\to e^\pm\mu^\mp X\,\,(X\to e^\pm\mu^\mp).
	\label{eq:associated-production-intro}
\end{equation}
Here we consider the subsequent decay $X\to e^\pm\mu^\mp$ and
retain the charge assignments of the four
final-state leptons. The signal can then be
separated into two categories:
\begin{align}
	\mathrm{CB}:&\qquad
	e^+e^-\mu^+\mu^-,
	\label{eq:CB-intro}\\
	\mathrm{SSSF}:&\qquad
	e^+e^+\mu^-\mu^-
	\quad\text{or}\quad
	e^-e^-\mu^+\mu^+.
	\label{eq:SSSF-intro}
\end{align}
The irreducible SM background in the CB channel is
\begin{equation}
	e^{+}e^{-}\to e^{+}e^{-}\mu^{+}\mu^{-}.
	\label{eq:bkgd}
\end{equation}
Its tree-level contributions
include neutral-current annihilation, radiation of virtual
photons or $Z$ bosons from charged-lepton lines, and
two-boson-fusion (multiperipheral) diagrams. All diagrams
and their interferences are included in its calculation.
Tau-decay backgrounds can also arise from
$e^+e^-\to\mu^+\mu^-\tau^+\tau^-$ with both taus decaying
to electrons, and from $e^+e^-\to e^+e^-\tau^+\tau^-$
with both taus decaying to muons. Both channels produce
the $e^+e^-\mu^+\mu^-$ charged-lepton configuration together
with four neutrinos. The corresponding double-decay
branching fractions are approximately 3.2\% and 3.0\%,
respectively, with a further
reduction from the charged-lepton selection requirements.
Four-muon production would require a double $\mu\to e$
misidentification and is negligible for particle-flow
detectors at future $e^+e^-$ colliders~\cite{Yu:2017mpx}.

In the SM, the SSSF final states arise from four-$\tau$
production, $e^+e^-\to\tau^+\tau^-\tau^+\tau^-$ (including the
$ZZ$ and $ZH$ contributions), followed by leptonic $\tau$ decays,
which is accompanied by eight neutrinos. The process $e^+e^-\to W^+W^-\tau^+\tau^-$, dominated by $ZH$ with
$H\to WW^*$ and $Z\to\tau^+\tau^-$, gives the same final state with
six neutrinos. A further
contribution arises from lepton charge misidentification in CB
events. Since the total charge of the four leptons is required to
vanish, this requires the simultaneous charge misidentification
of one electron and one muon.
\subsection{Event generation and common selection}
To execute the full simulation, we translate the effective interaction given in
Eq.~\eqref{eq:symmetric-interaction} into UFO model files using FeynRules2
(v2.3.49)~\cite{Alloul:2013bka}. Signal and background events are generated at
leading order with MadGraph5\_aMC@NLO (v3.5.6)~\cite{Alwall:2014hca}, showered with
PYTHIA~8 (v8316)~\cite{Sjostrand:2014zea}, which includes QED final-state radiation and
$\tau$ decays, and passed through DELPHES~3 (v3.5.0)~\cite{deFavereau:2013fsa} using the
standard FCC-ee card with Si tracking capabilities~\cite{FCC:2018evy,Suarez:2022pcn}.
The samples are generated for a nominal centre-of-mass energy
$\sqrt{s}=\sqrtsFCC$, with no beam-energy spread. They are normalised
to $\mathcal{L}=\lumiFCC$, the baseline integrated luminosity of the
$240~\mathrm{GeV}$ run with four interaction
points~\cite{FCC:2025lpp,Dam:2025zed}.

Signal events are generated for the charge combinations of
$e^+e^-\to e^\pm\mu^\mp X$ followed by $X\to e^\pm\mu^\mp$, in the narrow-width
approximation, for $\mx=$~10-220 GeV. The coupling is fixed at $h=0.01$, and the
width is computed for each mass from Eq.~\eqref{eq:X-width}. For $\mx=50$, $100$, and
$150~\GeV$ the signal cross sections summed over the CB and SSSF configuration are
$0.425$, $0.11$, and $0.0336$~fb, respectively, and the CB and SSSF cross sections agree
to within~0.5\%, as expected from Eq.~\eqref{eq:equal-charge-rates}.
The CB backgrounds $e^+e^-\to e^+e^-\mu^+\mu^-$, $e^+e^-\tau^+\tau^-$, and
$\mu^+\mu^-\tau^+\tau^-$ are generated together, with $X$ removed from all internal
propagators, giving a total cross section of $0.244$~pb. The four-$\tau$ background is
generated as a separate sample, with a cross section of 3.52~fb. Generator-level
requirements of $\pt^{\ell}>3$~GeV (with $\pt^{\ell_1}>5~\GeV$ for the leading
lepton), $|\eta_\ell|<$~3.0, and $\Delta R_{\ell\ell}>$~0.2 are applied to all
leptons, including the decay products of $X$; these are looser than the selection below.
We generate $5\times10^5$ events for each signal mass point, for the CB background, and
for the four-$\tau$ background, and normalise each sample with the weight
$w=\sigma\mathcal{L}/N_{\rm gen}$.

\par To study the kinematic distributions, we adopt selection criteria (acceptance
cuts) for the transverse momenta \((\pt^{\ell})\), pseudorapidities \((\eta_\ell)\), and
separations \((\Delta R_{\ell\ell})\) of the visible leptons \(\ell\equiv(e^{\pm},\,\mu^{\pm})\),
and for their total energy $E_{4\ell}$:
\begin{align}
 &\pt^{\ell_1}>10~\GeV,
 \qquad
 \pt^{\ell}>5~\GeV,
 \label{eq:pt-preselection}\\
 &|\eta_\ell|<2.5,
 \qquad
 \Delta R_{\ell\ell}>0.4,
 \qquad
 E_{4\ell}=\sum_{i=1}^{4}E_{\ell_i}>200~\GeV,
 \label{eq:eta-dr-preselection}
\end{align}
where $\ell_1$ denotes the leading charged lepton and the $5~\GeV$ threshold applies to
all four leptons. The energy requirement exploits the absence of invisible particles in
the signal, for which $E_{4\ell}\simeq\sqrt{s}$, and removes backgrounds with neutrinos,
in particular four-$\tau$ production.
We demand exactly two reconstructed electrons and two reconstructed muons in the final
state and the total charge is required to vanish. The category is then fixed by the
charges of the like-flavour pairs:
\begin{align}
	q(e_1)q(e_2)=-1,\quad q(\mu_1)q(\mu_2)=-1
	&\quad\Longrightarrow\quad \mathrm{CB},
	\label{eq:CB-classifier}\\
	q(e_1)q(e_2)=+1,\quad q(\mu_1)q(\mu_2)=+1
	&\quad\Longrightarrow\quad \mathrm{SSSF}.
	\label{eq:SSSF-classifier}
\end{align}
The two regions are mutually exclusive.

\subsection{Search strategy for charge-balanced channel}
The CB final state is given by
\begin{equation}
	e^+e^-\mu^+\mu^-.
\end{equation}
To analyze the invariant mass distributions of the flavor-violating charged lepton pairs, we define two opposite-sign, opposite-flavor invariant masses:
\begin{equation}
	m_1=m(e^+\mu^-),
	\qquad
	m_2=m(e^-\mu^+).
	\label{eq:CB-masses}
\end{equation}
In order to identify the correct pair we propose the following fixed mass hypothesis \(m_{ X}^{\rm test}\),
\begin{equation}
	m_{\rm best}^{\rm CB}=
	\begin{cases}
		m_1,
		& |m_1-\mx^{\rm test}|<|m_2-m_{ X}^{\rm test}|,\\[3pt]
		m_2,
		& |m_2-\mx^{\rm test}|<|m_1-m_{ X}^{\rm test}|,
	\end{cases}
	\label{eq:mbest-CB}
\end{equation}
where best-matched mass \(m_{\rm best}^{\rm CB}\) is the one closest to $m_{ X}^{\rm test}$.
\begin{figure}[h!]
  \centering
    \hspace{-1cm}\includegraphics[width=8.5cm,height=6cm]{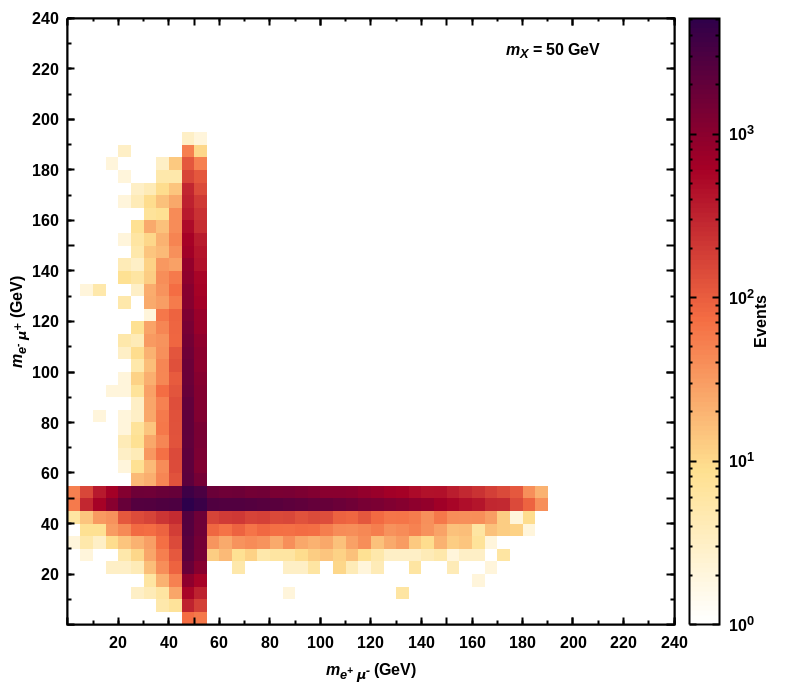}
    \includegraphics[width=7.5cm,height=6cm]{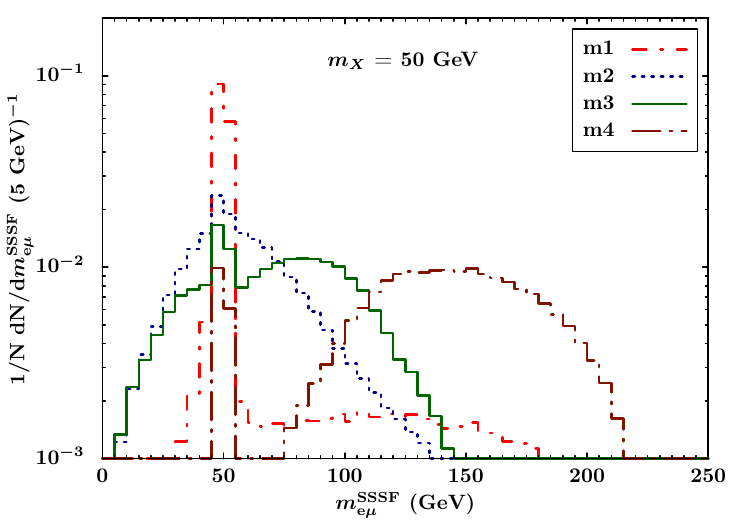}\\
    \hspace{-1cm}\includegraphics[width=8.5cm,height=6cm]{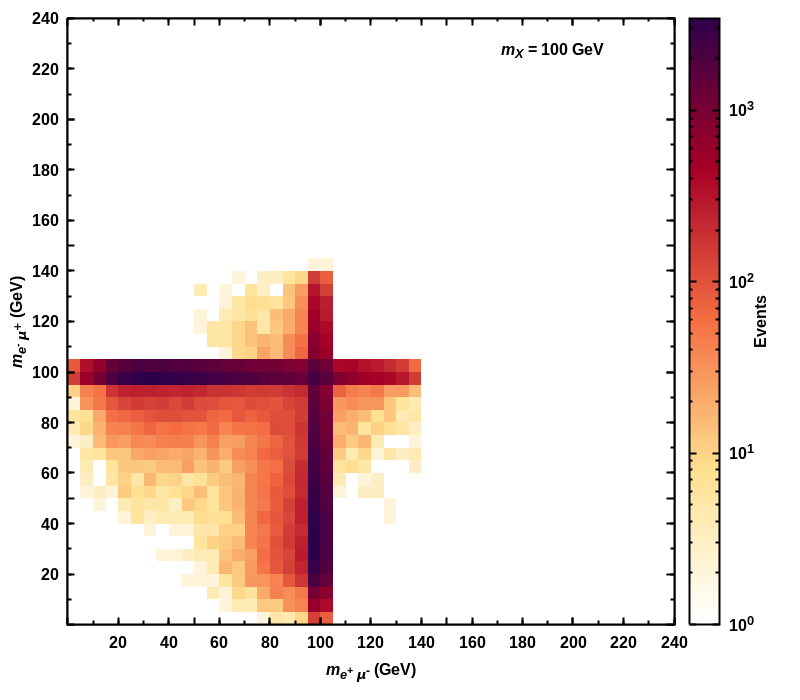}
    \includegraphics[width=7.5cm,height=6cm]{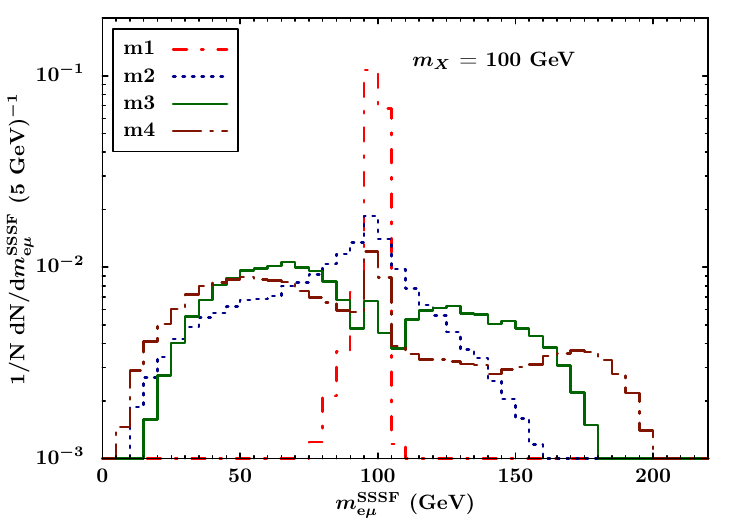}\\
    \hspace{-1cm}\includegraphics[width=8.45cm,height=6cm]{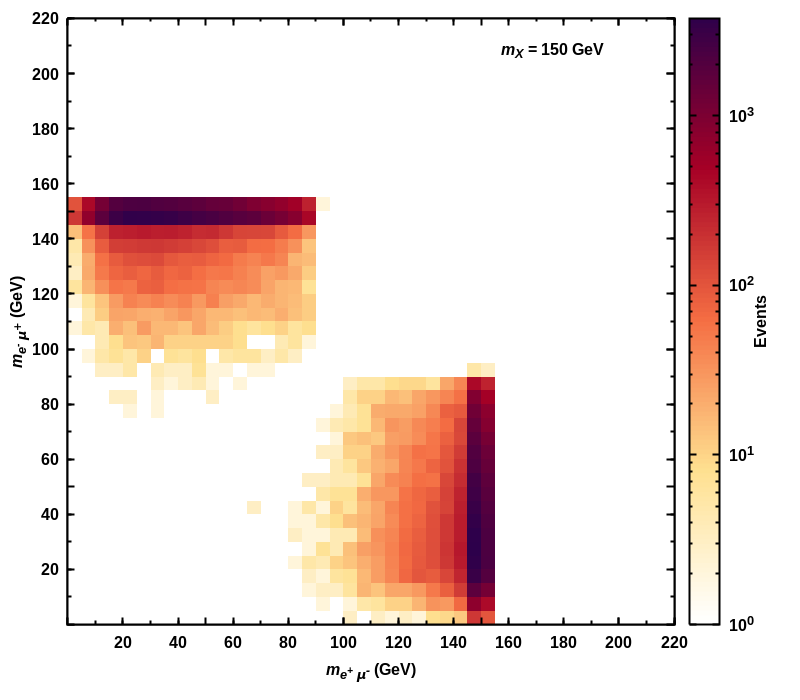}
    \includegraphics[width=7.5cm,height=6cm]{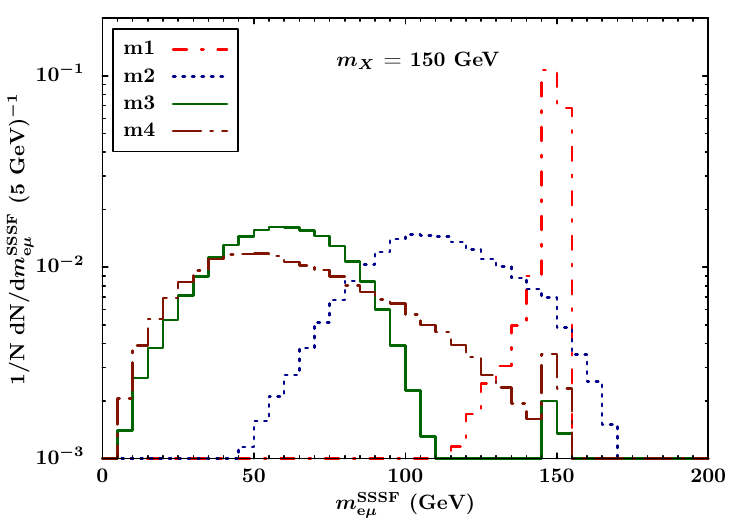}
  \caption{Mass reconstruction in the CB channel (left) and the SSSF
  channel (right). The CB plane shows both opposite-sign $e\mu$ masses,
  while the SSSF variable is selected from the four allowed $e\mu$
  combinations.}
  \label{fig:mass-reconstruction}
\end{figure}

\par To investigate, we show both $m_1$ and $m_2$, or equivalently the two-dimensional distribution of the opposite-sign, opposite-flavour invariant masses, $m_{e^+\mu^-}$ and $m_{e^-\mu^+}$. The reconstruction of the scalar resonance in the CB channel is illustrated in the left panel of Fig.~\ref{fig:mass-reconstruction} for three representative masses, $m_{ X}=50$, $100$, and $150$ GeV. The distributions exhibit characteristic horizontal and vertical bands near the corresponding scalar mass, reflecting the two possible charge assignments of the $e\mu$ pair originating from the scalar decay. For $m_X=50$ GeV, both structures are prominent, with the nonresonant invariant mass extending over a broad kinematic range. As $m_X$ increases, the two bands become more distinctly separated, with the $100$ and $150$ GeV benchmarks showing pronounced concentrations around the respective resonance masses. These features in the CB channel motivate the mass-based strategy adopted later in the analysis.

\subsection{Search strategy for same-sign same-flavour channel}
\label{subsec:SSSF}

The SSSF final states are given by
\begin{equation}
	e^+e^+\mu^-\mu^-
	\qquad\text{and}\qquad
	e^-e^-\mu^+\mu^+.
	\label{eq:SSSF-finalstates}
\end{equation}
Although the two electrons and the two muons have equal charges within each flavour, every candidate \( X\) decay remains an opposite-sign, opposite-flavour pair. This leads to four distinct invariant-mass combinations.

For $e^+e^+\mu^-\mu^-$, the four combinations are defined as
\begin{equation}
	m_{ij} = m(e_i^+\mu_j^-),
	\qquad i,j=1,2.
	\label{eq:SSSF-four-masses}
\end{equation}
The reconstructed candidate is selected according to
\begin{equation}
	m_{\mathrm{best}}^{\mathrm{SSSF}}
	= \operatorname*{arg\,min}_{m_{ij}}
	\bigl| m_{ij} - m_{{X}}^{\mathrm{test}} \bigr|,
	\label{eq:mbest-SSSF}
\end{equation}
Therefore, $m_{\rm best}^{\rm SSSF}$ is the invariant mass with the smallest deviation from $m_{{X}}^{\mathrm{test}}$. The charge-conjugate category is treated in the same way.

The right panel of Fig.~\ref{fig:mass-reconstruction} shows the
normalized invariant-mass distributions for the four opposite-sign,
opposite-flavour pairings in the SSSF channel. For each event, the
combinations are ordered by their absolute difference from
$m_X^{\mathrm{test}}$ and labelled $m_1$ through $m_4$, where
$m_1$ ($m_4$) has the smallest (largest) difference. We therefore
identify $m_1$ as the best-matched candidate
$m_{\rm best}^{\rm SSSF}$. Its distribution exhibits a clear peak
near $m_X$, while the alternative combinations populate broader
mass regions. For $m_X=50$ GeV, the distributions overlap
substantially near the resonance. The best-matched peak becomes
more clearly separated from the alternative combinations for the
heavier benchmarks, particularly at $m_X=150$ GeV.

\par
This mass dependence can be understood qualitatively from the
associated-production topology in Fig.~\ref{fig:signal-diagrams},
where the scalar is emitted from a charged-lepton line through
the flavour-changing $Xe\mu$ vertex. At fixed collision energy,
lighter scalars can be more strongly boosted. The resulting
changes in the energy sharing and angular separation of the
decay leptons affect the invariant masses formed with the
associated leptons and hence the overlap among the candidate
distributions. For heavier scalars, less energy is available to
the associated leptons, which can help separate the resonance
from the alternative combinations. The prominence of the $m_1$
peak also reflects the ordering prescription, which selects the
mass closest to the test hypothesis by construction.

\par
These distributions illustrate the resonance reconstruction.
The SSSF sensitivity is obtained using only the common
preselection, without an additional invariant-mass window or
mass-dependent transverse-momentum requirement.

\par
There is no SM process producing the final states in
Eq.~\eqref{eq:SSSF-finalstates} without additional neutrinos.
These configurations change the electron and muon flavour numbers
by two units, whereas the SM conserves them separately when
neutrino-mass-induced flavour violation is neglected. The SSSF background therefore arises from four-$\tau$ and $W^+W^-\tau^+\tau^-$ production followed by leptonic $\tau$ and $W$ decays, which are accompanied by neutrinos and suppressed by the $E_{4\ell}$ requirement in Eq.~\eqref{eq:eta-dr-preselection}, and from lepton charge misidentification in CB events. Because the total lepton charge
is required to vanish, the latter requires the simultaneous charge
misidentification of one electron and one muon. These contributions
are quantified in Sec.~\ref{subsec:reach}.

\section{Results and discussions}
\label{sec:results}
\subsection{Kinematic behaviour and cut flow}

The signal contains four visible charged leptons and no invisible particle.
For events without hard unrecovered radiation,
\begin{equation}
  E_{4\ell}=\sum_{i=1}^{4}E_{\ell_i}\simeq\sqrt{s},
  \qquad
  \left|\sum_{i=1}^{4}\vec p_{T,\ell_i}\right|\simeq0.
  \label{eq:exclusive-kinematics}
\end{equation}
The first property is used in the preselection,
Eq.~\eqref{eq:eta-dr-preselection}, to remove backgrounds with neutrinos.
The direct invariant mass of the correct $e\mu$ pair is the primary
resonance observable. The recoil mass does not contain independent
information when all four leptons are reconstructed; the proof is given in
Appendix~\ref{app:recoil}.

\begin{figure}[h!]
  \centering
  \includegraphics[width=4.5cm,height=4cm]{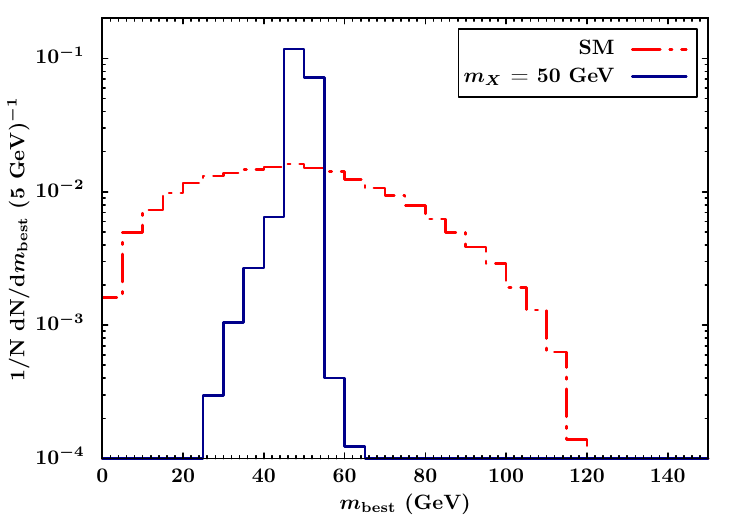}
  \includegraphics[width=4.5cm,height=4cm]{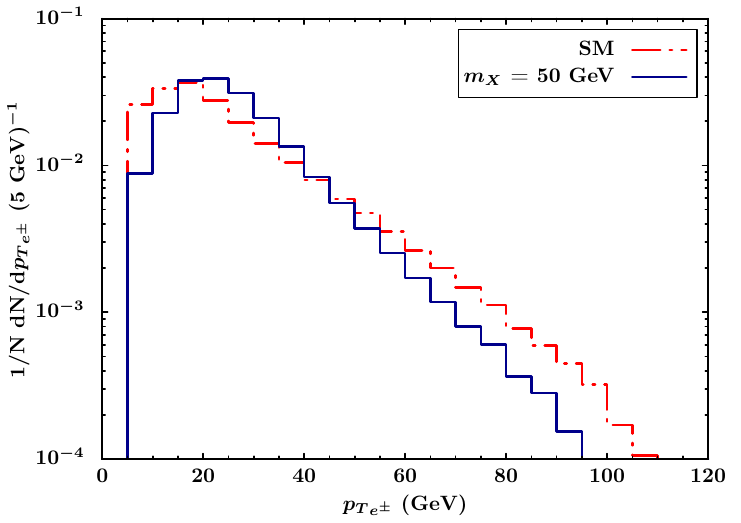}
  \includegraphics[width=4.5cm,height=4cm]{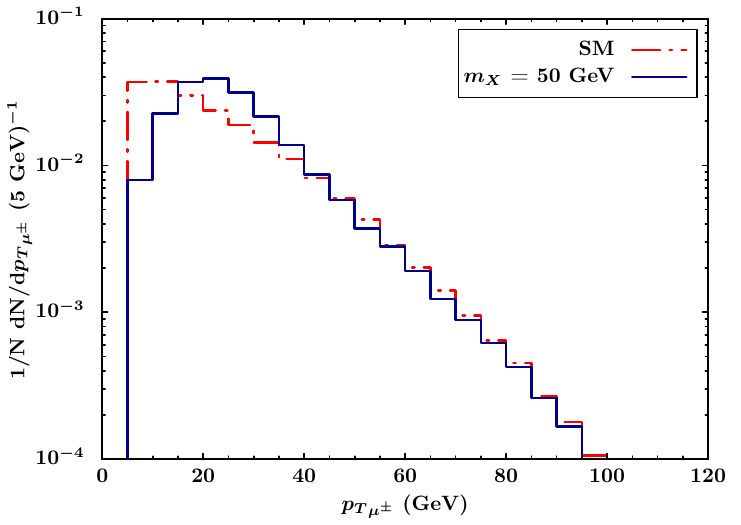}\\
  \includegraphics[width=4.5cm,height=4cm]{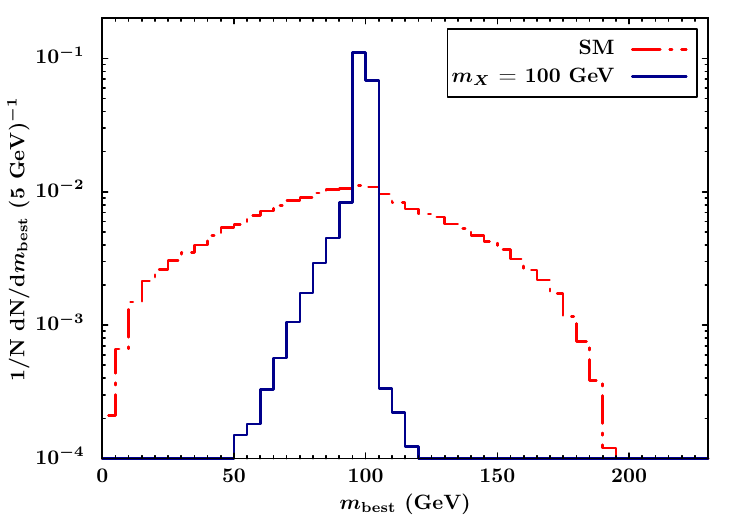}
  \includegraphics[width=4.5cm,height=4cm]{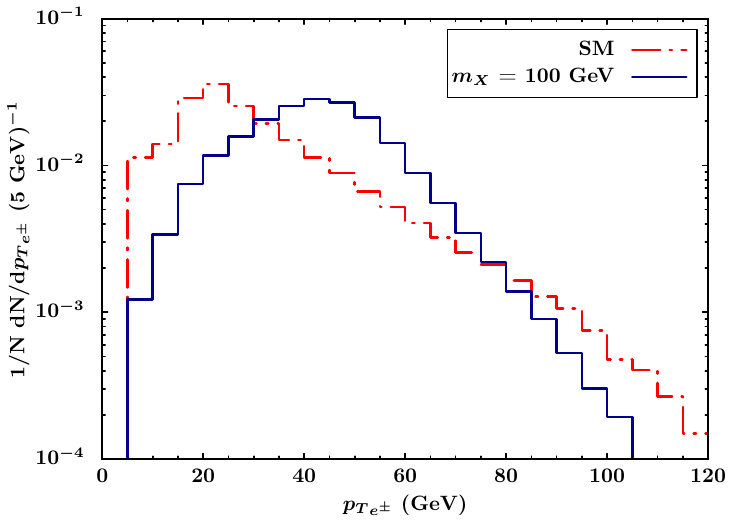}
  \includegraphics[width=4.5cm,height=4cm]{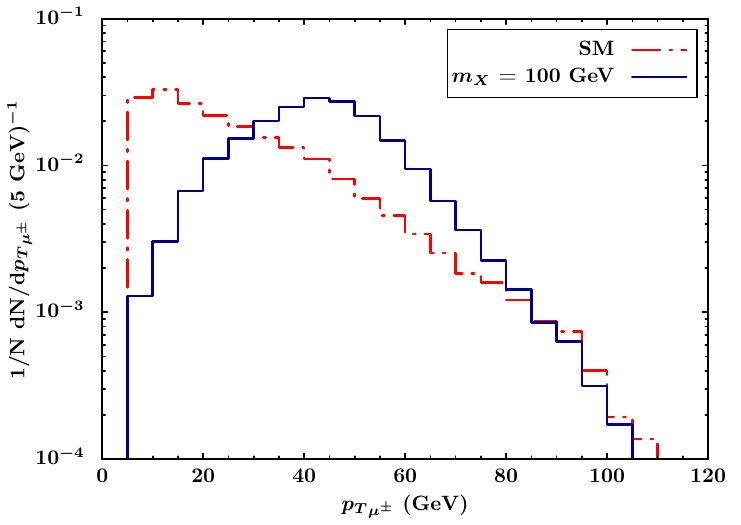}\\
  \includegraphics[width=4.5cm,height=4cm]{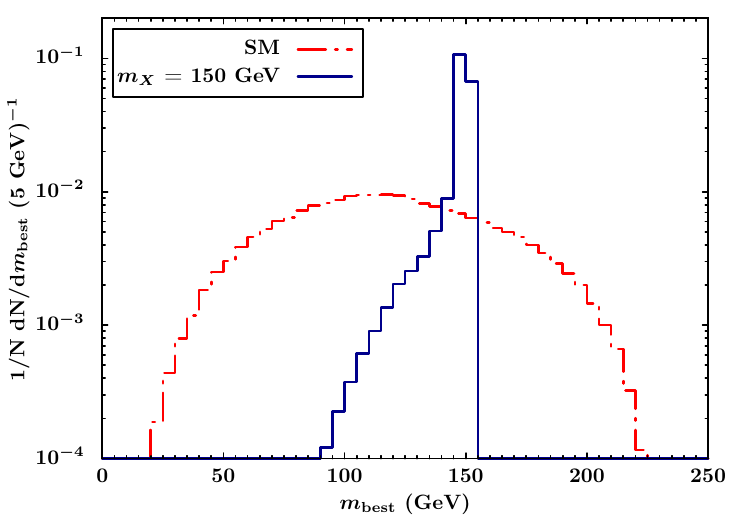}
  \includegraphics[width=4.5cm,height=4cm]{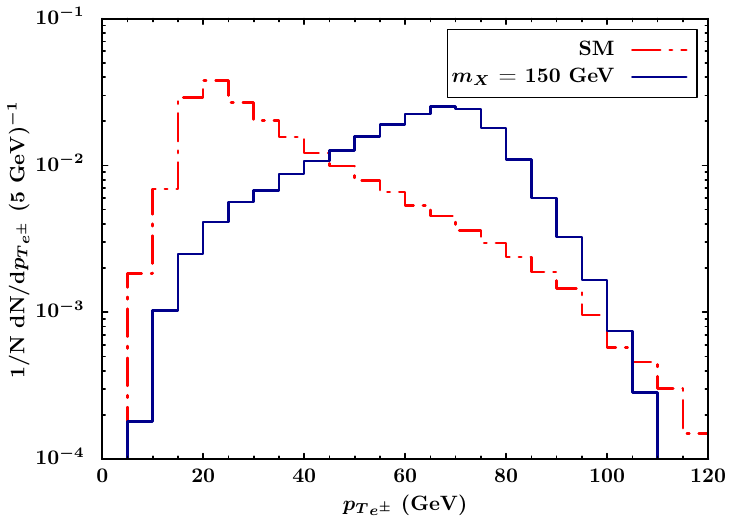}
  \includegraphics[width=4.5cm,height=4cm]{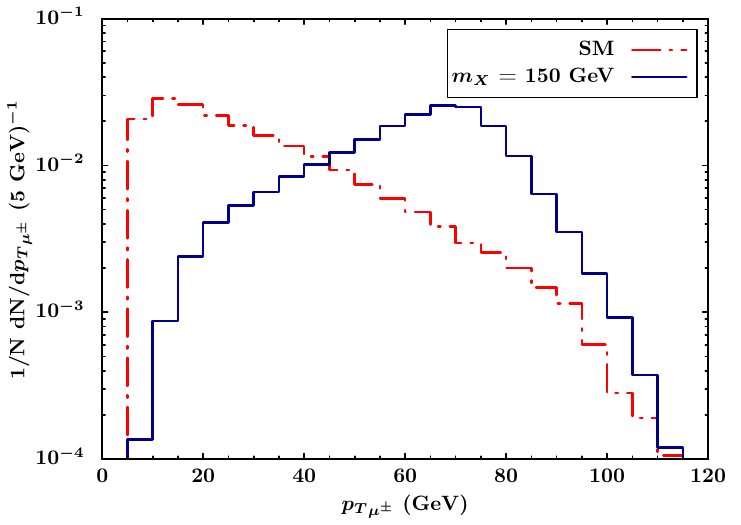}\\  \caption{Kinematic distributions for the charge-balanced (CB) channel at $\sqrt{s}=240~\mathrm{GeV}$, comparing the signal (solid blue) with the SM background (red dash-dotted) for $m_X=50$, $100$, and $150~\mathrm{GeV}$ (top to bottom). The columns show the reconstructed invariant mass $m_{\mathrm{best}}$, electron transverse momentum, and muon transverse momentum, respectively. All distributions are normalized to unit area.}
  \label{fig:distr_cb}
\end{figure}

\begin{figure}[h!]
  \centering
  \includegraphics[width=4.5cm,height=4cm]{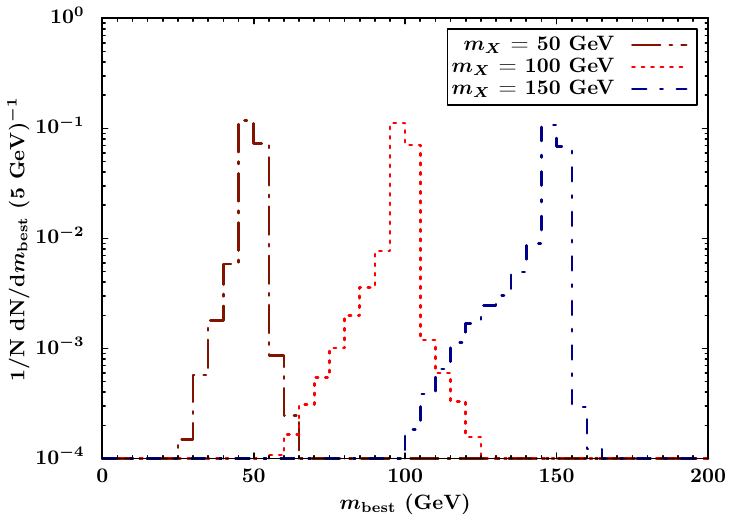}
  \includegraphics[width=4.5cm,height=4cm]{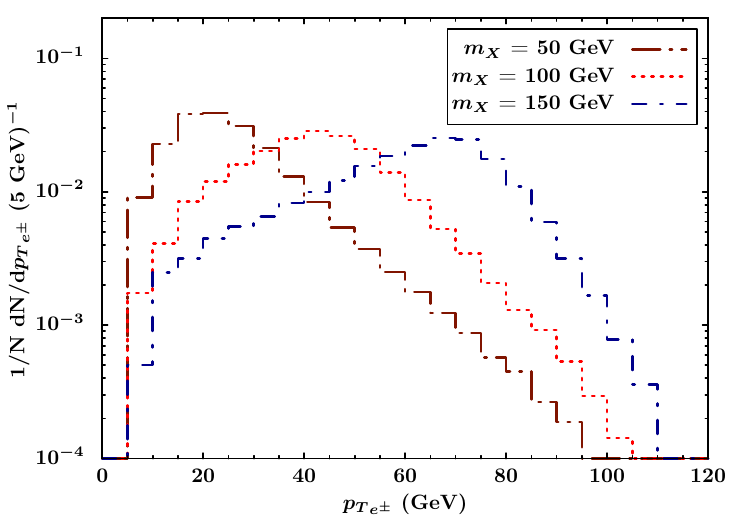}
  \includegraphics[width=4.5cm,height=4cm]{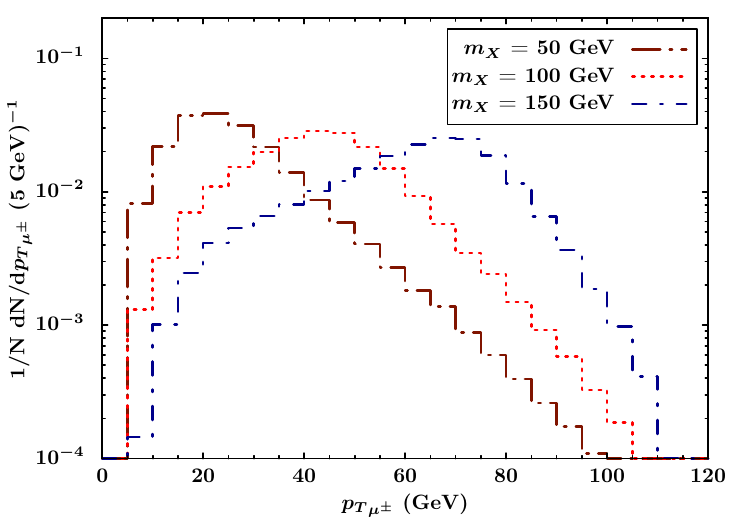}\\  \caption{Normalized kinematic distributions for the same-sign same-flavour (SSSF) channel at $\sqrt{s}=240~\mathrm{GeV}$. From left to right, the panels show the reconstructed invariant mass $m_{\mathrm{best}}$, electron transverse momentum, and muon transverse momentum. The three scalar mass benchmarks, $m_X=50$, $100$, and $150~\mathrm{GeV}$, are distinguished by different line styles. All distributions are normalized to unit area.}
  \label{fig:distr_sssf}
\end{figure}

The normalized kinematic distributions for the charge-balanced (CB) channel are presented in Fig.~\ref{fig:distr_cb} for three representative scalar masses, $m_X=50$, $100$, and $150$ GeV. The reconstructed invariant mass $m_{\mathrm{best}}$ exhibits a pronounced peak near the corresponding scalar mass, while the SM background follows a broader, nonresonant distribution. The separation becomes more pronounced for the heavier benchmarks. The electron and muon transverse-momentum distributions also show a clear dependence on the scalar mass. As $m_X$ increases, both spectra shift towards higher transverse momenta, reflecting the harder leptons produced in the decay of a heavier scalar. In comparison, the SM background is concentrated at lower $p_T$, making these observables useful for further signal discrimination.

The reconstructed peak is narrow: its width is set by the detector
resolution and final-state radiation, since the intrinsic width
$\Gamma_X/m_X=h^2/4\pi$ is negligible. We therefore require
\begin{equation}
  |m_{\rm best}^{\rm CB}-\mx^{\rm test}|<\Delta m_X,
  \qquad
  \Delta m_X=10~\GeV,
  \label{eq:CB-mass-window}
\end{equation}
which is wide compared with the reconstructed peak and retains
\(\sim\)~92-98\% of the preselected signal for $m_X=50$--$150~\GeV$.
To further suppress the SM background, we require both leptons of the best-matched pair
to satisfy $\pt^{\ell}>m_X/3$. These additional invariant-mass and transverse-momentum
requirements are applied only to the CB channel.

The corresponding distributions for the same-sign same-flavour (SSSF) channel are shown in Fig.~\ref{fig:distr_sssf}. The reconstructed invariant-mass peaks remain well defined for all three benchmarks, with their positions closely following $m_X$. The transverse-momentum spectra exhibit a similar mass dependence to those in the CB channel, becoming progressively harder as the scalar mass increases.

\subsection{Projected FCC-ee sensitivity}
\label{subsec:reach}

We now analyze the projected sensitivity of FCC-ee to the
lepton-flavour-violating coupling $h$ at
$\sqrt{s}=\sqrtsFCC$, assuming an integrated luminosity
of $\mathcal{L}=\lumiFCC$. The sensitivity is evaluated
separately for the CB and SSSF final states, accounting
for their different background conditions.

\begin{table}[h!]
  \centering
  \caption{Cut flow for the charge-balanced (CB) channel at $h=0.01$. Entries give the
  cross section in fb after each selection, with the efficiency relative to the four-lepton
  preselection in parentheses, for the signal benchmarks $m_X=50$, $100$, and $150~\GeV$
  and the corresponding SM background. Each sample contains $5\times10^5$ generated
  events; the expected number of events follows from multiplying by
  $\mathcal{L}=\lumiFCC$.}
  \label{tab:CB-cutflow}
  \resizebox{\textwidth}{!}{%
  \begin{tabular}{lrrrrrr}
    \toprule
    & \multicolumn{3}{c}{Signal} & \multicolumn{3}{c}{CB background} \\
    \cmidrule(lr){2-4}\cmidrule(lr){5-7}
    Selection
      & $50~\GeV$ & $100~\GeV$ & $150~\GeV$
      & $50~\GeV$ & $100~\GeV$ & $150~\GeV$ \\
    \midrule
    Four-lepton preselection
      & 0.0980 (100\%) & 0.0283 (100\%) & 0.0086 (100\%)
      & 28.12 (100\%) & 28.12 (100\%) & 28.12 (100\%)\\
    Mass window, Eq.~\eqref{eq:CB-mass-window}
      & 0.0959 (97.86\%)& 0.0266 (93.99\%) & 0.0079 (91.86\%) & 8.55 (30.40\%) & 5.94 (21.12\%)& 3.71 (13.19\%)\\
    $\pt^{\ell}>m_X/3$
      & 0.0612 (62.45\%) & 0.0162 (57.24\%) & 0.0052 (60.46\%)& 3.74 (13.30\%)& 1.12 (3.98\%)& 0.47 (1.67\%)\\
    \bottomrule
  \end{tabular}%
  }
\end{table}

\begin{table}[h!]
  \centering
  \caption{Expected SSSF yields after the full SSSF selection, including
  $E_{4\ell}>200~\GeV$, at $\mathcal{L}=\lumiFCC$. The signal is shown for
  $h=0.01$. The four-$\tau$ entry is a 95\% CL upper limit, since no simulated
  event passes the selection. The $W^+W^-\tau^+\tau^-$ entry is the estimate described in the text. For charge misidentification, the value in
  parentheses corresponds to the conservative rates $\varepsilon_e=10^{-2}$
  and $\varepsilon_\mu=10^{-3}$.}
  \label{tab:SSSF-yields}
  \begin{tabular}{lc}
    \toprule
    Process & Expected events \\
    \midrule
    Signal, $\mx=50~\GeV$  & $1060$ \\
    Signal, $\mx=100~\GeV$ & $304$ \\
    Signal, $\mx=150~\GeV$ & $93$ \\
    \midrule
    $e^+e^-\to\tau^+\tau^-\tau^+\tau^-$   & $<0.23$ \\
    $e^+e^-\to W^+W^-\tau^+\tau^-$   & $\lesssim0.2$ \\
    Charge misidentification in CB events & $0.06$ $(6)$ \\
    \midrule
    Total background                      & $<0.5$ $(6.5)$ \\
    \bottomrule
  \end{tabular}
\end{table}

The effectiveness of the event selection in the CB channel is summarized in
Table~\ref{tab:CB-cutflow}. Following preselection, we impose the mass window of
Eq.~\eqref{eq:CB-mass-window} around the reconstructed scalar resonance and require
both leptons of the best-matched pair to satisfy $\pt^{\ell}>m_X/3$. The mass window
retains~92-98\% of the signal while reducing the SM background to~13-30\%, depending on
the scalar mass hypothesis. The subsequent transverse-momentum requirement yields
cumulative background efficiencies of~1.7-13\%, while retaining~57-62\% of the
preselected signal. The SSSF yields are summarized in Table~\ref{tab:SSSF-yields}.

The CB sensitivity is evaluated from a binned likelihood analysis of
$m_{\rm best}^{\rm CB}$ in $0.5~\GeV$ bins inside the mass window of
Eq.~\eqref{eq:CB-mass-window}, after all CB selections, at
$\mathcal{L}=\lumiFCC$. The signal yield in bin $i$ is simulated at $h=0.01$
and rescaled as $S_i(m_X,h)=S_i(m_X,0.01)\,(h/0.01)^2$, which holds since the
scalar remains narrow, $\Gamma_X/m_X=h^2/4\pi\lesssim2\times10^{-3}$, over the
range shown. Including a background uncertainty $\Delta_i=\delta_B B_i$ with
$\delta_B=1\%$, uncorrelated between bins, we use the median exclusion
significance~[Eq.~(1.6) of Ref.~\cite{Kumar:2015tna}]
\begin{equation}
  Z_i = \Big[ 2\Big( S_i - B_i\ln\frac{B_i+S_i+x_i}{2B_i}
     - \frac{B_i^2}{\Delta_i^2}\ln\frac{B_i-S_i+x_i}{2B_i}\Big)
     - (B_i+S_i-x_i)\Big(1+\frac{B_i}{\Delta_i^2}\Big)\Big]^{1/2},
  \label{eq:Z-CB}
\end{equation}
with $x_i=\sqrt{(S_i+B_i)^2-4S_iB_i\Delta_i^2/(B_i+\Delta_i^2)}$, which reduces to
$S_i/\sqrt{B_i+\Delta_i^2}$ for $S_i\ll B_i$. The bins are combined as
$Z_{\rm CB}=\big(\sum_i Z_i^2\big)^{1/2}$, and the expected 95\% CL exclusion
contour is defined by $Z_{\rm CB}=1.645$. The result is insensitive to the
window size: varying it between $\pm5$ and $\pm20~\GeV$ changes the reach by
less than $1\%$. With $1~\GeV$ bins the reach is weaker by about
$15$--$20\%$, and with $5~\GeV$ bins by a factor of about two.

In the SSSF channel, the four-$\tau$ background is estimated from
$5\times10^5$ simulated events with
$\sigma(e^+e^-\to\tau^+\tau^-\tau^+\tau^-)=3.52$~fb. Without the $E_{4\ell}$
requirement, $206$ simulated events pass the SSSF selection, corresponding to
about $16$ expected events; with it, none pass, corresponding to
$b_{4\tau}<3\,\sigma\mathcal{L}/N_{\rm gen}=0.23$ events at 95\% CL. The $W^+W^-\tau^+\tau^-$ background, dominated by $ZH$ with $H\to WW^*$
and $Z\to\tau^+\tau^-$, gives about 15 SSSF events before any selection.
Its six neutrinos typically carry about half of the collision energy,
whereas $E_{4\ell}>200~\GeV$ allows them at most $40~\GeV$ in total.
For four-$\tau$ production this requirement removes all 206 simulated
events that pass the other SSSF requirements, corresponding to an
efficiency below $1.5\%$ at 95\% CL. With the assumed efficiency, we estimate about \(0.2\) $W^+W^-\tau^+\tau^-$ events. The charge-misidentification
contribution is
\begin{equation}
  b_{\rm flip}\simeq 2\,\varepsilon_e\varepsilon_\mu\,N_{\rm CB},
  \label{eq:charge-flip}
\end{equation}
where $\varepsilon_{e}$ ($\varepsilon_{\mu}$) is the probability to
misidentify the charge of an electron (muon), $N_{\rm CB}\simeq3.0\times10^5$
is the expected number of CB background events after the common
preselection (Table~\ref{tab:CB-cutflow}), and the factor of two accounts
for the two charge-conjugate configurations. We take
$\varepsilon_e=10^{-3}$ and $\varepsilon_\mu=10^{-4}$ as benchmark values.
These lie within the ranges measured at the LHC, where the electron charge
misidentification probability varies from $2\times10^{-5}$ for low-$\pt$
central electrons to $10^{-2}$ at high $\pt$ near the edge of the barrel
calorimeter, reaching $10\%$ at larger $|\eta|$~\cite{ATLAS:2018alq}, and
the muon charge misassignment rises from below $10^{-4}$ at $10~\GeV$ to
about $10^{-2}$ at $500~\GeV$~\cite{CMS:2009fdy}. Since the lepton momenta
at FCC-ee are below about $120~\GeV$, we also consider
$\varepsilon_e=10^{-2}$ and $\varepsilon_\mu=10^{-3}$ as a conservative
variation. This gives $b_{\rm flip}\simeq0.06$ ($6$) events for the
benchmark (conservative) rates.

For such small backgrounds the asymptotic expression in Eq.~\eqref{eq:Z-CB},
derived in the large-count limit, overestimates the significance; for
$b\to0$ it reduces to $Z=\sqrt{2s}$, which would imply an exclusion with
$s\simeq1.4$ events. We therefore use the exact Asimov significance of
Ref.~\cite{Bhattiprolu:2020mwi}, obtained from the Poisson $p$-value with the
observed count replaced by its background-only expectation,
\begin{equation}
  p_{\rm excl}=\frac{\Gamma(b+1,\,s+b)}{\Gamma(b+1)},
  \label{eq:exact-asimov}
\end{equation}
where $\Gamma(a,x)$ is the upper incomplete gamma function. An expected
95\% CL exclusion, $p_{\rm excl}=0.05$, then requires $s_{95}(b)$ selected
signal events, with $s_{95}=\ln 20\simeq2.996$ for $b\to0$, $3.09$ for
$b=0.1$, $3.74$ for $b=1$, and $5.51$ for $b=5$. The SSSF reach is
\begin{equation}
  h_{\rm SSSF}=0.01\,\sqrt{\frac{s_{95}(b)}{S_{\rm SSSF}(\mx,0.01)}},
  \label{eq:SSSF-reach}
\end{equation}
where $S_{\rm SSSF}(\mx,0.01)$ is the expected number of selected SSSF
signal events at $h=0.01$ and $\mathcal{L}=\lumiFCC$, and $b$ is the total
SSSF background. For the nominal background estimate in Table~3 ($b\lesssim 0.5$),
we use the low-background approximation $s_{95}\simeq 3$ for the
SSSF curves in Figs.~5 and~6. Evaluating Eq.~(5.5) at $b=0.5$
increases the projected upper limit on $|h|$ by about $7\%$.
For the conservative charge-misidentification rates,
$s_{95}\simeq 6.0$, which weakens the coupling reach by a factor
of about $1.4$.

\begin{figure}[t]
  \centering
  \includegraphics[width=12 cm, height=10 cm]{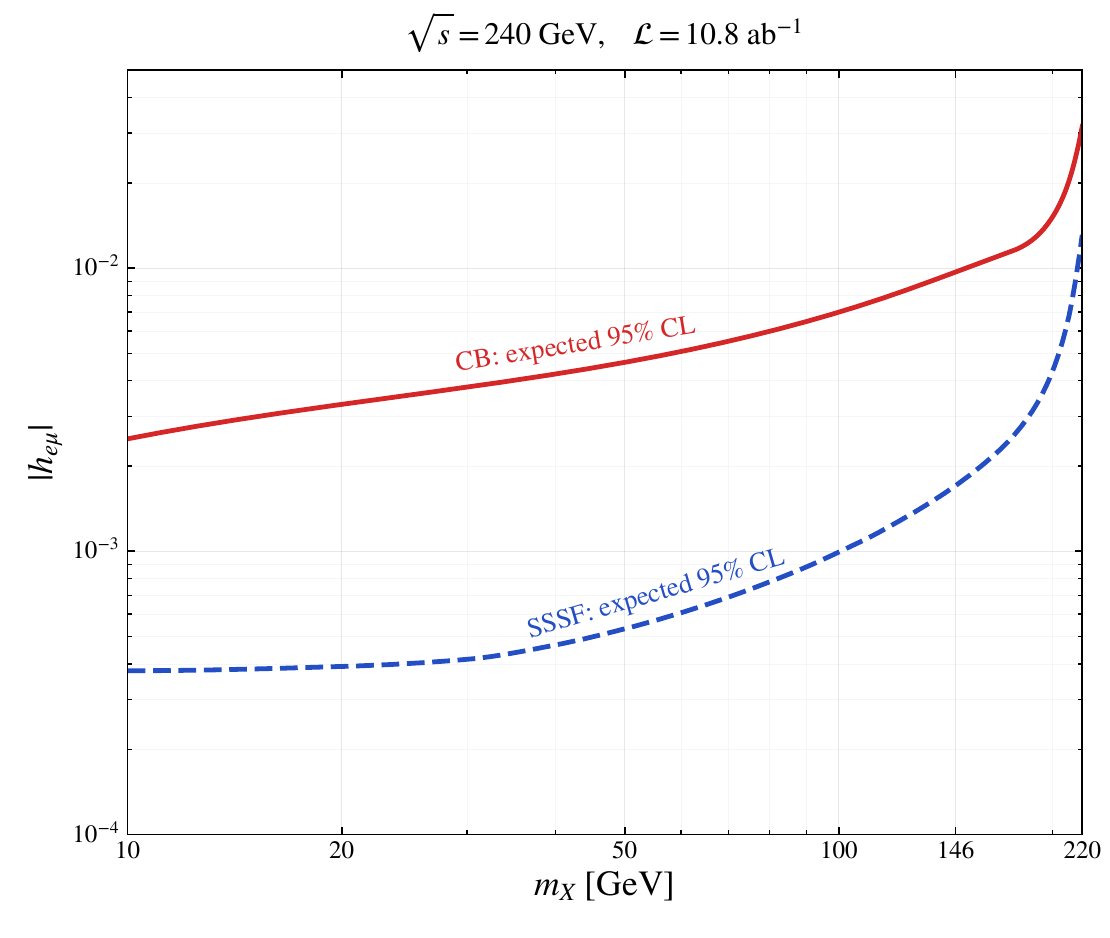}
  \caption{FCC-ee projections for the $e\mu$ coupling at $\sqrt{s}=\sqrtsFCC$ and
  $\mathcal{L}=\lumiFCC$. Both curves show the expected 95\% CL exclusion: CB from
  Eq.~\eqref{eq:Z-CB} with a 1\% background uncertainty, and SSSF from
  Eq.~\eqref{eq:SSSF-reach}.}
  \label{fig:reach-only}
\end{figure}

The resulting expected 95\% CL contours are shown in Fig.~\ref{fig:reach-only}.
Both channels exhibit their strongest sensitivity in the
low-mass region, with the reach deteriorating as the scalar
mass increases. The degradation becomes particularly
pronounced near the kinematic limit, where the available
phase space for associated production is strongly
suppressed. Throughout the mass range considered, the
SSSF curve lies below the CB contour, indicating access
to smaller LFV couplings. Since the two charge categories
have equal signal rates before charge-asymmetric acceptance
effects, Eq.~\eqref{eq:equal-charge-rates}, this
difference is primarily associated with their background
conditions and with the additional CB selection.

At the CMS benchmark mass, the expected 95\% CL reaches are
$|h|_{\rm CB}\simeq\hCBcms$ and $|h|_{\rm SSSF}\simeq\hSSSFcms$.
Our analysis incorporates parton showering and detector simulation within the
FCC-ee environment for the reconstruction and selection of final-state leptons,
and includes the SSSF background estimates in Table 3. Fake leptons from hadron decays and photon conversions are suppressed by
lepton identification at future $e^+e^-$ detectors~\cite{Yu:2017mpx}, by lepton
isolation, by the requirement of exactly four leptons with vanishing total
charge, and by the $E_{4\ell}$ requirement, and are not simulated.

\subsection{Comparison with present bounds}

\begin{figure}[t]
  \centering
  \includegraphics[width=0.80\textwidth]{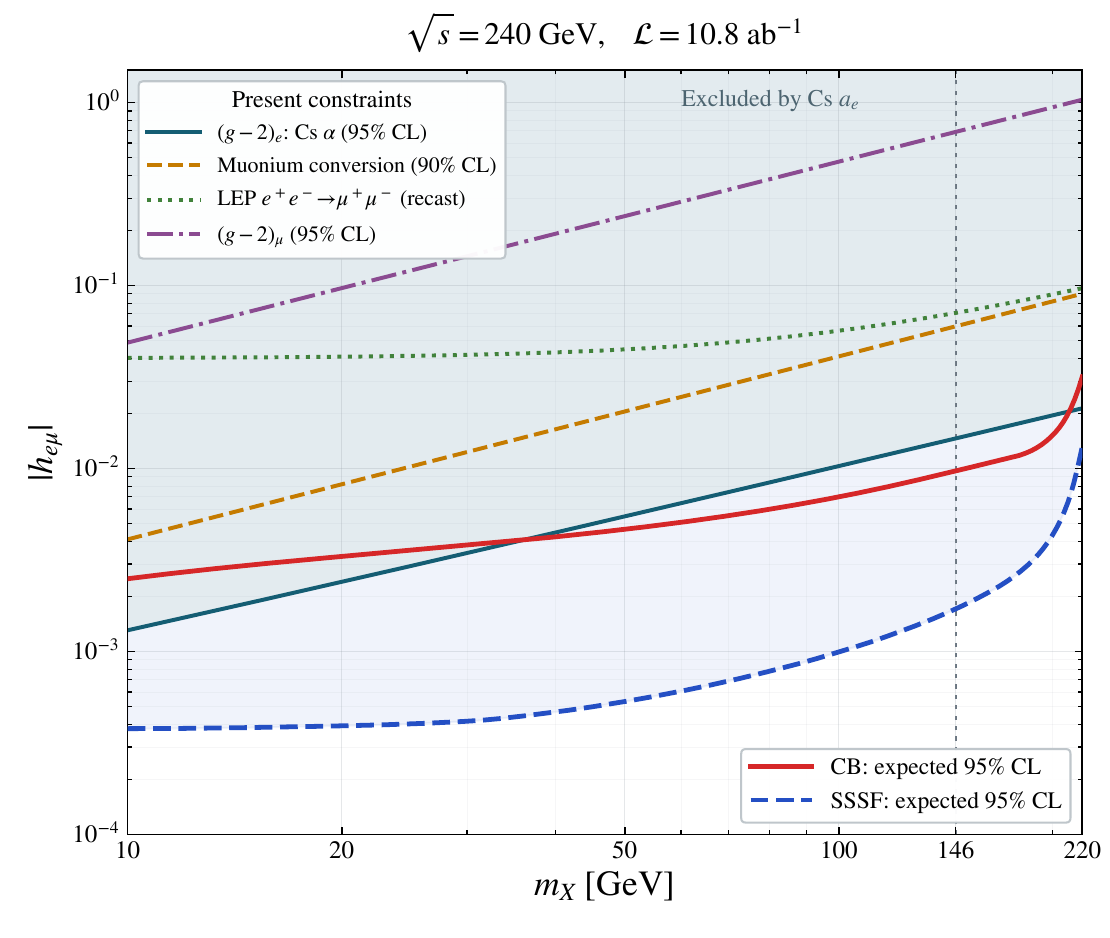}
  \caption{Present one-coupling constraints and FCC-ee sensitivities in the
$(\mx,|h|)$ plane. The magnetic-moment bounds are shown at 95\% C.L. and the muonium bound at 90\% C.L., as published; the LEP curve is the recast
described in Sec.~\ref{sec:constraints}. The shaded region above the
electron-$g-2$ curve is excluded for the cesium input and a positive CP-even scalar contribution; the shaded band between the electron-$g-2$ curve and the SSSF curve is accessible to FCC-ee but not excluded by present bounds. The CB and SSSF curves show the expected 95\% CL exclusion at $\mathcal{L}=\lumiFCC$. The vertical line marks $\mx=146~\GeV$.}
  \label{fig:constraints-and-reach}
\end{figure}
We now compare the projected FCC-ee sensitivity with the
existing constraints on the LFV coupling in the one-coupling
scenario. Figure~\ref{fig:constraints-and-reach} summarizes
the bounds from the electron and muon anomalous magnetic
moments, muonium-antimuonium conversion, and an approximate
recast of LEP searches for
$e^+e^-\to\mu^+\mu^-$, together with the CB and SSSF
sensitivity contours derived above.

Among the constraints considered, the electron anomalous
magnetic moment provides the strongest bound over most of
the mass range shown. The corresponding exclusion is
obtained using the cesium determination of the fine-structure
constant and the positive contribution of the CP-even
scalar to $a_e$. The muonium-conversion upper bound on $|h|$ increases with $m_X$ and remains weaker than the electron-$g-2$ bound in the mass range shown. The LEP constraint provides an
independent restriction on the LFV interaction, while
the muon anomalous magnetic moment is considerably less
restrictive in this parameter space. The CB sensitivity
is weaker than the electron-$g-2$ constraint for $m_X\lesssim35~\GeV$ and
extends beyond it for $35\lesssim m_X\lesssim210~\GeV$. The SSSF contour falls below the existing exclusion boundary over the whole mass range up to about $220~\GeV$, opening a region of smaller LFV
couplings to direct collider investigation.

At $m_X=146\,\mathrm{GeV}$, the expected 95\% CL CB reach is
$|h|_{\rm CB}\simeq\hCBcms$, compared with the electron-$g-2$ bound
$|h|_{(g-2)_e}\simeq\haecms$. The expected 95\% CL SSSF reach is
$|h|_{\rm SSSF}\simeq\hSSSFcms$, a factor of about 9 below this bound.

\section{Conclusions}
\label{sec:conclusions}

We have investigated the collider phenomenology of a real
CP-even neutral scalar $X$ with a single symmetric
lepton-flavour-violating $e\mu$ coupling at FCC-ee.
In this minimal framework, the same interaction governs
the associated production
$e^+e^-\to e^\pm\mu^\mp X$
and the subsequent decay
$X\to e^\pm\mu^\mp$, giving rise to distinctive
four-lepton signatures.

We have considered two mutually exclusive charge
configurations: the charge-balanced (CB) final state
$e^+e^-\mu^+\mu^-$ and the same-sign same-flavour (SSSF)
final states $e^\pm e^\pm\mu^\mp\mu^\mp$.
The CB channel receives an irreducible SM four-lepton
continuum background, whereas the SSSF topology arises in the SM only from processes with neutrinos, mainly four-$\tau$ and $W^+W^-\tau^+\tau^-$ production, and from lepton charge misidentification, both of which we find
to be small after the selection.
The reconstructed $e\mu$ invariant-mass distributions
exhibit narrow resonant structures near $m_X$, with
improved separation between the resonant and alternative
lepton pairings for the heavier scalar benchmarks.
The invariant-mass and lepton transverse-momentum
distributions provide the principal observables for
signal discrimination. Following the event selection,
the CB channel retains~57-62\%
of the preselected signal events for
$m_X=50$--$150~\GeV$, while reducing the corresponding
background fractions to~1.7-13\%, respectively.

At $\sqrt{s}=\sqrtsFCC$ and an integrated luminosity
of $\mathcal{L}=\lumiFCC$, we have evaluated the
projected sensitivity in the $(m_X,|h|)$ plane.
The analysis incorporates parton showering and detector
simulation within the FCC-ee environment, including
the reconstruction and selection of the final-state
charged leptons. The CB sensitivity is obtained from a
binned analysis of the reconstructed $e\mu$ mass,
including a 1\% background uncertainty, and the SSSF
sensitivity from a counting analysis including the using its background estimates;
both are expressed as expected 95\% CL exclusions.
The SSSF channel
probes smaller LFV couplings throughout the
investigated mass range, primarily owing to the
different background conditions of the two charge
categories. Both sensitivities deteriorate as
$m_X$ approaches the kinematic limit, where the
available phase space for associated scalar
production becomes increasingly restricted.

We have compared these projections with the
constraints from the electron and muon anomalous
magnetic moments, muonium–antimuonium conversion,
and LEP dilepton measurements. At the CMS-motivated
benchmark mass $m_X=146~\GeV$, the CB channel
reaches $|h|\simeq\hCBcms$, while the
SSSF channel probes couplings down to
$|h|\simeq\hSSSFcms$. The CB channel also extends beyond the electron-$g-2$ bound for $35\lesssim m_X\lesssim210~\GeV$.
The SSSF channel lies a factor of about 9
below the cesium-based electron-$g-2$ bound,
$|h|\simeq\haecms$, demonstrating the
potential of the SSSF signature to access
parameter space beyond the existing constraints
considered here.

The SSSF four-lepton final state provides a promising
signature for investigating electron--muon flavour violation at future
electron--positron colliders. The present projection includes the simulated
detector response, the four-$\tau$ and $WW\tau\tau$ backgrounds, and lepton charge
misidentification. A full experimental study would additionally include fake leptons from hadron decays, and photon
conversions with a dedicated detector simulation.

\section*{Acknowledgements}
The authors thank Debajyoti Choudhury for useful discussions. KD is supported by the Croatian Science Foundation under the
project number HRZZ-IP-2025-02-4414.
TS is supported by the DST, Government of India, through the DST INSPIRE Faculty Fellowship
(DST/INSPIRE/04/2024/004616). This work is supported by IoE, University of Delhi project Grant no. IoE/2025-2026/12/FRP, which provided the computational resources used to perform the simulations in this study.
\appendix

\section{Recoil mass and invariant-mass reconstruction}
\label{app:recoil}

Let the two opposite-sign $e\mu$ systems in a CB event be
\begin{equation}
  q_1=p_{e^+}+p_{\mu^-},
  \qquad
  q_2=p_{e^-}+p_{\mu^+}.
\end{equation}
For an initial four-momentum
\begin{equation}
  P_0=(\sqrt{s},\vec 0),
\end{equation}
the recoil mass against $q_i$ is
\begin{equation}
  m_{{\rm rec},i}^2
  =(P_0-q_i)^2
  =s+m_i^2-2\sqrt{s}\,E_i.
  \label{eq:recoil-mass}
\end{equation}
If all four leptons are reconstructed and there is no unrecovered radiation,
\begin{equation}
  P_0=q_1+q_2.
\end{equation}
It follows immediately that
\begin{equation}
  m_{{\rm rec},1}^2=q_2^2=m_2^2,
  \qquad
  m_{{\rm rec},2}^2=q_1^2=m_1^2.
  \label{eq:recoil-equals-opposite-mass}
\end{equation}
Choosing the recoil mass closest to a benchmark mass is therefore equivalent
to choosing the opposite-pair invariant mass closest to that benchmark. ISR,
beam-energy spread, and lost photons break the equality, but they generally
broaden the recoil mass while leaving the direct invariant mass of the
$X$ decay products less affected. We therefore use the $e\mu$ invariant mass
as the primary resonance variable.

\bibliographystyle{JHEP}
\bibliography{ref}

\end{document}